# *The Parabolic-Trigonometric Functions*

*G. Dattoli, M. Migliorati*[*]*, M. Quattromini and P. E. Ricci*[†]

ENEA, Tecnologie Fisiche e Nuovi Materiali, Centro Ricerche Frascati
C.P. 65 – 00044 Frascati, Rome, Italy

**ABSTRACT**

The parabolic functions are introduced in analogy to the circular and hyperbolic cases. We discuss the relevant properties, the geometrical interpretation and touch on possible generalizations and their link with the modular elliptic functions.

[*] Dipartimento di Energetica , Università di Roma “La Sapienza”, Via A. Scarpa 14 - 00161 Rome, Italy

[†] Dipartimento di Matematica, Università di Roma “La Sapienza”, P.le Aldo Moro, 5 – 00185 Rome, Italy

## 1. INTRODUCTION

This paper is aimed at filling a gap in the theory of elementary functions, we will introduce indeed the trigonometric parabolic function and study the relevant properties.

The circular and hyperbolic functions are well known, the associated geometrical interpretation is provided in Figs. 1,2, while the geometric images of what we mean by parabolic-trigonometric functions are shown in Fig. 3.

Before getting into more technical aspects and in order to clarify the reasons which inspired the present research, we note that it does not appear as widespread known, as it should be, the fact that the argument of the hyperbolic functions is just the area of the sector indicated in Fig. 2.

The hyperbolic functions can be therefore viewed as parametric functions, through the parameter $\Phi$.

The link between hyperbolic and trigonometric functions is usually done using the imaginary unit, but it will be more useful for our purposes, the introduction of the so called Gudemann function $gd(\Phi)$ [1], whose role can be understood as follows.

According to Fig. 2 the relation between the parameter $\Phi$ and the corresponding angle $\varphi$ is provided by

$$tg(\varphi) = tgh(\Phi) \tag{1a}$$

so that the Gudermann function writes‡

$$\varphi = gd(\Phi) = tg^{-1}\left[tgh(\Phi)\right] \tag{1b}$$

‡ The link between hyperbolic and circular functions is ensured by identities of the type

$$\cos(gd(\Phi)) = \frac{\cosh(\Phi)}{\sqrt{\sinh(\Phi)^2 + \cosh(\Phi)^2}}$$

by taking the derivative with respect to $\Phi$, we obtain the relation

$$\frac{d}{d\Phi} gd(\Phi) = \sec h(2\Phi) \tag{2}$$

which yields the alternative definition

$$gd(\Phi) = \int_0^{\Phi} \sec h(\sigma) d\sigma \tag{3}.$$

We can take advantage from the above definition to get a deeper insight in the theory of generalized trigonometric functions. Ferrari proposed in ref. (2) the following definition (n is any positive integer)

$$C(\Phi \mid n)^n + S(\Phi \mid n)^n = 1, \tag{4}$$

with the parameter $\Phi$ specified by

$$\frac{1}{2} C(\Phi \mid n) S(\Phi \mid n) + \int_{C(\Phi|n)}^{1} \left[1 - \xi^n\right]^{\frac{1}{n}} d\xi = \frac{1}{2}\Phi \tag{5}$$

In Fig. (4) we have reported the case $n = 3$, along with the relevant geometrical insight.

The equations (4,5) can be exploited to state the relevant properties under derivation

$$\begin{aligned} &\frac{d}{d\Phi} C(\Phi \mid n) = -S(\Phi \mid n)^{n-1}, \\ &\frac{d}{d\Phi} S(\Phi \mid n) = C(\Phi \mid n)^{n-1} \end{aligned} \tag{6a}.$$

Which, along with the conditions

$$C(0 \mid n) = 1, \; S(0 \mid n) = 0 \tag{6b}$$

can be exploited to derive the associated series expansions, as more carefully discussed below.

We will consider the following generalization of eqs. (4,5), with $p,q$ two distinct integers

$$\begin{aligned} & C(\Phi \mid p,q)^p + S(\Phi \mid p,q)^q = 1, \\ & \frac{1}{2} C(\Phi \mid p,q) S(\Phi \mid p,q) + \int_{C(\Phi|p,q)}^{1} \left[1-\xi^p\right]^{\frac{1}{q}} d\xi = \frac{1}{2}\Phi \end{aligned} \tag{7}$$

which, once combined, yield the derivation rules

$$\begin{aligned} & \frac{d}{d\Phi} C(\Phi \mid p,q) = -q \frac{S(\Phi \mid p,q)^{q-1}}{qS(\Phi \mid p,q)^q + pC(\Phi \mid p,q)^p}, \\ & \frac{d}{d\Phi} S(\Phi \mid p,q) = p \frac{C(\Phi \mid p,q)^{p-1}}{qS(\Phi \mid p,q)^q + pC(\Phi \mid p,q)^p} \end{aligned} \tag{8}$$

It is also proved by direct check that

$$\begin{aligned} & \frac{d}{d\Phi} T(\Phi \mid p,q) = \frac{1}{C(\Phi \mid p,q)^2}, \\ & T(\Phi \mid p,q) = \frac{S(\Phi \mid p,q)}{C(\Phi \mid p,q)} \end{aligned} \tag{9}$$

In the following we will specialize the previous general results to particular values of the integers $p,q$.

## 2. The Parabolic Functions: Elementary properties

We will identify the parabolic trigonometric functions with

$$\begin{aligned} & C(\Phi \mid 2,1) = \cos p(\Phi), \\ & S(\Phi \mid 2,1) = \sin p(\Phi), \\ & \cos p(0) = 1, \sin p(0) = 0 \end{aligned} \tag{10}$$

they satisfy the fundamental relation

$$\cos p(\Phi)^2 + \sin p(\Phi) = 1 \tag{11}$$

whose geometrical interpretation is given in Figs. 3 and the relevant derivatives write

$$\begin{aligned} &\frac{d}{d\Phi}\cos p(\Phi) = -\frac{1}{\sin p(\Phi) + 2\cos p(\Phi)^2}, \\ &\frac{d}{d\Phi}\sin p(\Phi) = 2\frac{\cos p(\Phi)}{\sin p(\Phi) + 2\cos p(\Phi)^2} \end{aligned} \qquad (12).$$

The first of eqs. (12) and eq. (11) yields the following differential equation, defining the parabolic cosine

$$\frac{d}{d\Phi}Y_P(\Phi) + \frac{1}{3}\frac{d}{d\Phi}Y_P(\Phi)^3 = -1 \qquad (13),$$

which on account of the fact that $\cos p(0) = 1$ can be exploited to derive the algebraic equation

$$Y_P(\Phi)^3 + 3Y_P(\Phi) + 3\Phi - 4 = 0 \qquad (14),$$

whose solution defines explicitly the parabolic cosine in terms of irrational functions, namely

$$\cos p(x) = \sqrt[3]{\frac{4-3x}{2} + \sqrt{1 + \frac{(4-3x)^2}{4}}} - \sqrt[3]{\sqrt{1 + \frac{(4-3x)^2}{4}} - \frac{4-3x}{2}} \qquad (15a).$$

The parabolic sine, defined through the fundamental identity (11), writes

$$\sin p(x) = -\sqrt[3]{\left[\frac{4-3x}{2} + \sqrt{1 + \frac{(4-3x)^2}{4}}\right]^2} - \sqrt[3]{\left[\sqrt{1 + \frac{(4-3x)^2}{4}} - \frac{4-3x}{2}\right]^2} + 3 \qquad (15b)$$

the relevant plot is shown in Fig. 5, in the interval

$$0 \le \Phi \le \frac{8}{3} \tag{16}$$

whose upper limit will be clarified below.

The series expansion of the parabolic trigonometric functions can either be obtained from eq. (15) or from the definitions (10-12), thus getting

$$\begin{aligned} \cos p(\Phi) &\cong 1 - \frac{\Phi}{2} - \frac{\Phi^2}{8} - \frac{\Phi^3}{24} - \frac{1}{384}(5\Phi^4 + \Phi^5) \\ \sin p(\Phi) &\cong \Phi - \frac{\Phi^3}{24} - \frac{\Phi^4}{32} - \frac{7}{384}\Phi^5 \end{aligned} \qquad (17).$$

The Gudermann parabolic function plays the same role as for the hyperbolic case and can be written as

$$\mathrm{gdp}(\Phi) = \mathrm{tg}^{-1}\left[\mathrm{tgp}(\Phi)\right] \tag{18}$$

and the relevant behaviour is shown in Fig. 6

The parameter $\Phi^*$ corresponding to $\cos p(\Phi^*) = 0, \sin p(\Phi^*) = 1$ is evidently given by

$$\frac{\Phi^*}{2} = \int_0^1 (1 - \xi^2) d\xi = \frac{2}{3} \tag{19}$$

in correspondence of which the Gudermann function yields

$$\mathrm{gdp}(\Phi^*) = \frac{\pi}{2} \qquad (20),$$

it is also evident that $\cos p(2\Phi^*) = -1, \sin p(2\Phi^*) = 0)$ and this clarifies the upper limit in eq. (16).

### 3. The link with the circular functions.

In the previous section we have shown that the definition of the parabolic functions can be traced back to a cubic algebraic solution, whose solution in terms of circular functions (or better ordinary transcendent functions) is well known (see the concluding section).

In this section we will discuss the link between parabolic and circular functions in a fairly detailed way.

The derivative of the parabolic Gudermann function reads

$$\frac{d}{d\Phi}\mathrm{gdP}(\Phi)=\frac{1}{[\mathrm{Ip}(\Phi)]^2} \tag{21}$$

where the parabolic secant writes (see Fig. 3a for further comments)

$$\mathrm{Ip}(\Phi)=\sqrt{\cos\mathrm{p}(\Phi)^2+\sin\mathrm{p}(\Phi)^2} \qquad (22).$$

The role of the function $\mathrm{Ip}(\Phi)$ is clear, at least from the geometrical point of view. According to Fig. 7 the ordinary theorem of rectangle triangles should, indeed, read as follows

$$\begin{aligned} C&=I\frac{\cos\mathrm{p}(\Phi)}{\mathrm{Ip}(\Phi)},\\ c&=I\frac{\sin\mathrm{p}(\Phi)}{\mathrm{Ip}(\Phi)} \end{aligned} \qquad (23).$$

Where $c, C, I$ denotes the edges of a rectangle triangle.

Let us now concentrate on the functions

$$\begin{aligned} \mathrm{cp}(\Phi)&=\frac{\cos\mathrm{p}(\Phi)}{\mathrm{Ip}(\Phi)}=\frac{1}{\sqrt{1+\mathrm{tgp}(\Phi)^2}}\\ \mathrm{sp}(\Phi)&=\frac{\sin\mathrm{p}(\Phi)}{\mathrm{Ip}(\Phi)}=\frac{\mathrm{tgp}(\Phi)}{\sqrt{1+\mathrm{tgp}(\Phi)^2}} \end{aligned} \qquad (24),$$

and note that they satisfy the following properties under derivation

$$\begin{aligned} &\mathrm{Ip}^2(\Phi)\frac{d}{d\Phi}\mathrm{cp}(\Phi) = -\mathrm{sp}(\Phi), \\ &\mathrm{Ip}^2(\Phi)\frac{d}{d\Phi}\mathrm{sp}(\Phi) = \mathrm{cp}(\Phi) \end{aligned} \tag{25}$$

Introducing moreover the function

$$\mathrm{Ep}(\Phi) = \mathrm{cp}(\Phi) + i\,\mathrm{sp}(\Phi) \tag{26}$$

we obtain, by differentiation, the identity

$$\mathrm{Ip}^2(\Phi)\frac{d}{d\Phi}\mathrm{Ep}(\Phi) = i\,\mathrm{Ep}(\Phi) \tag{27}$$

which once integrated yields

$$\mathrm{Ep}(\Phi) = e^{i\int_0^{\Phi}\frac{d\sigma}{\mathrm{Ip}^2(\sigma)}} \tag{28}.$$

The previous results are interesting for at least two reasons.

a) The functions $cp(\Phi), \sin p(\Phi)$ satisfy the fundamental identity

$$\mathrm{cp}(\Phi)^2 + \mathrm{sp}(\Phi)^2 = 1 \tag{29}$$

and behave, under the “derivative”

$$\hat{D} = I_p^2(\Phi)\frac{d}{d\Phi} \tag{30}$$

as the ordinary circular functions (see Eq. (25))

b) The function (28) is an eigen-function of the operator (30) and can be exploited to define the parabolic functions as

$$\begin{aligned} &\cos p(\Phi) = \mathrm{Ip}(\Phi)\cos\left(\int_0^{\Phi}\frac{d\sigma}{\mathrm{Ip}^2(\sigma)}\right), \\ &\sin p(\Phi) = \mathrm{Ip}(\Phi)\sin\left(\int_0^{\Phi}\frac{d\sigma}{\mathrm{Ip}^2(\sigma)}\right) \end{aligned} \tag{31}.$$

The previous conclusions are not limited to the parabolic function and can be extended to all the generalized trigonometric functions, as it will be shown in a forthcoming paper.

**4. Concluding Remarks**

In this paper we have outlined the properties of the parabolic functions, which seems to be ideally suited to treat oscillations ruled by the differential equation

$$\begin{aligned} & A(\Phi)\frac{d^2}{d\Phi^2}y + B(\Phi)\frac{d}{d\Phi}y + y = 0, \\ & B(\Phi) = \frac{1}{2}\frac{d}{d\Phi}A(\Phi) \end{aligned} \qquad (32),$$

whose solution can be written in terms of the linear combination

$$y(\Phi) = \left[\alpha\cos\left(\int_0^{\Phi}\frac{d\sigma}{\sqrt{A(\sigma)}}\right) + \beta\sin\left(\int_0^{\Phi}\frac{d\sigma}{\sqrt{A(\sigma)}}\right)\right] \qquad (33).$$

A further link of the parabolic trigonometric functions with the ordinary transcendent functions can be obtained by going back to the solution of the cubic equation, whose solution in terms of hyperbolic function provides the following identification for the parabolic functions

$$\begin{aligned} & \cos p(\Phi) = -2\sinh\left(\frac{1}{3}\sinh^{-1}\left(\frac{y(\Phi)}{2}\right)\right), \\ & \sin p(\Phi) = 3 - 2\cosh\left(\frac{2}{3}\sinh^{-1}\left(\frac{y(\Phi)}{2}\right)\right) \\ & y(\Phi) = 3\Phi - 4 \end{aligned} \qquad (34).$$

Before closing the paper it is worth stressing the above result, which opens new research lines in the field of elliptic functions.

The investigations of the properties of the generalized trigonometric functions may be indeed particularly interesting for their important relations with the Jacobi and Weierstrass functions, this relationship has been thoroughly studied by Ferrari, for the cases $n = 3, 5, 7$ of the $C(\Phi, n), S(\Phi, n)$ [2].

Here we want to emphasize that the functions

$$\begin{aligned} & C(\Phi \mid 4,1) = \cos m(\Phi), \\ & S(\Phi \mid 4,1) = \sin m(\Phi), \\ & \cos m(0) = 1, \sin m(0) = 0 \end{aligned} \qquad (35),$$

which can be defined through the algebraic quintic equation

$$3 \cos m(\Phi)^5 + 5 \cos m(\Phi) - 8 + 5\Phi = 0 \qquad (36)$$

may offer new tools in the theory of the solution of the fifth degree equations and may provide an alternative point of view on the theory of the modular elliptic functions [3], as it will be shown in a forthcoming investigation.

**ACKNOWLEDGMENTS**

One of the Authors (G. D.) expresses his sincere appreciation to the late Prof. Ezio Ferrari for introducing him to the theory of generalized Trigonometric Functions.

## REFERENCES

1. M. Abramowitz and I. Stegun “Handbook of Mathematical Functions” Dover Pub. New York (1964)

2. E. Ferrari, Bollettino UMI 18B, 933 (1981)

   For early suggestions see also F.D. Bugogne, Math. of Comp. 18, 314 (1964)

3. H. T. Davis “Introduction to non linear Differential and Integral Equations” Dover pub. New York (1962).

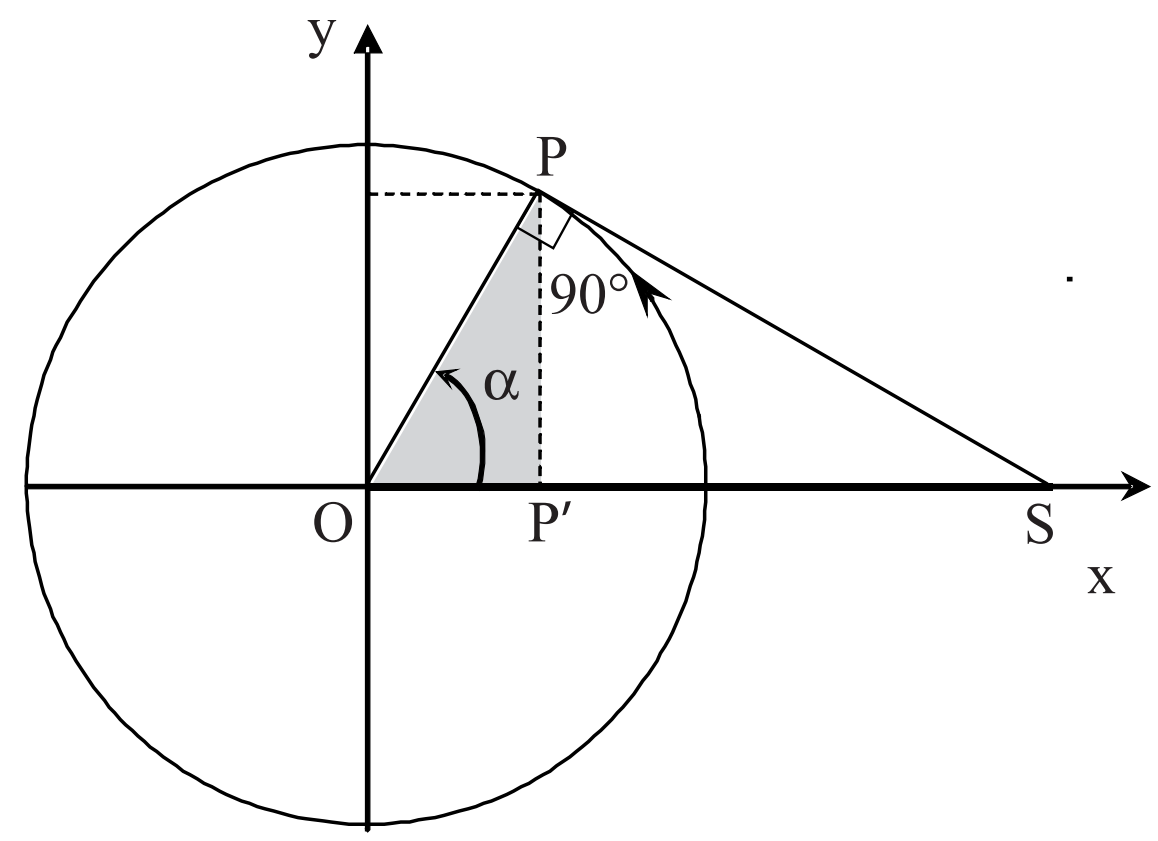

Fig. 1

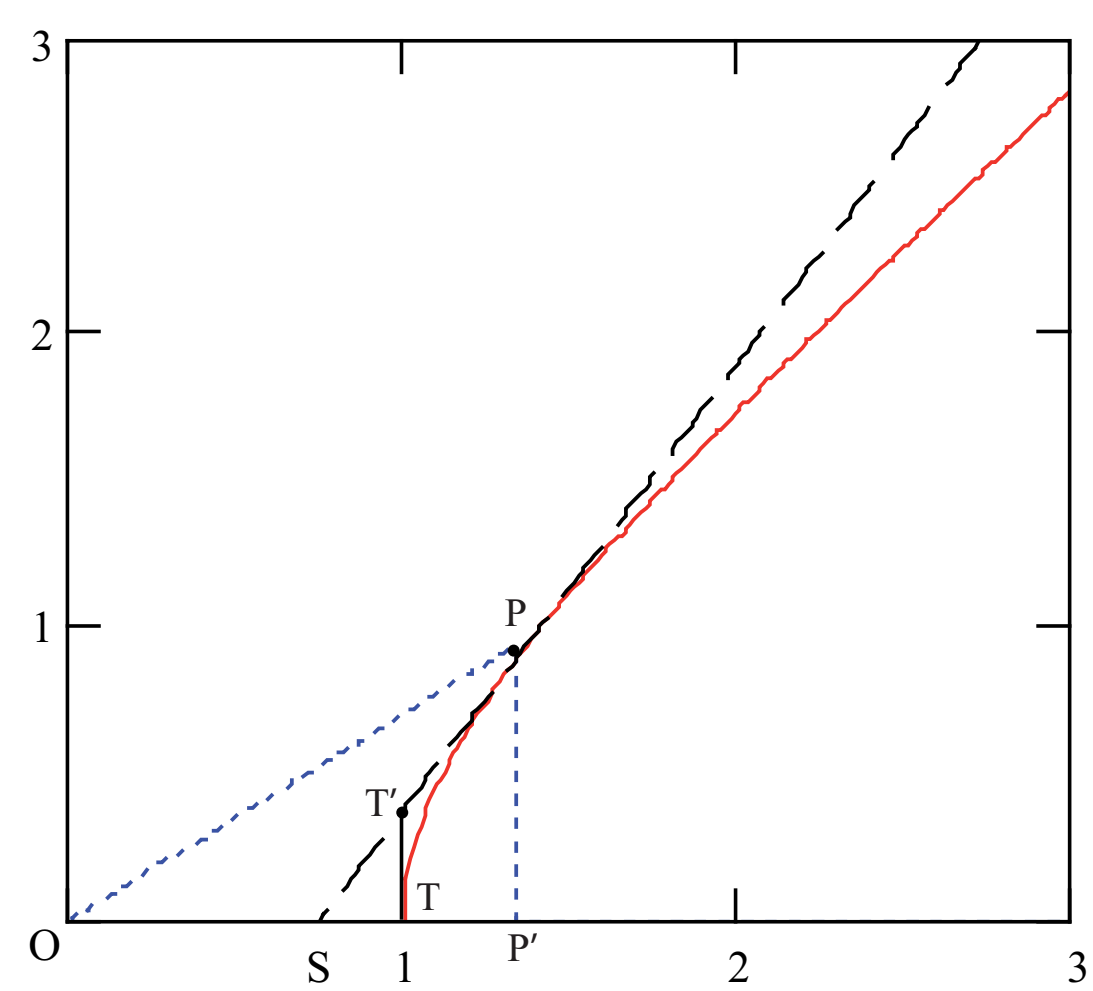

Fig. 2

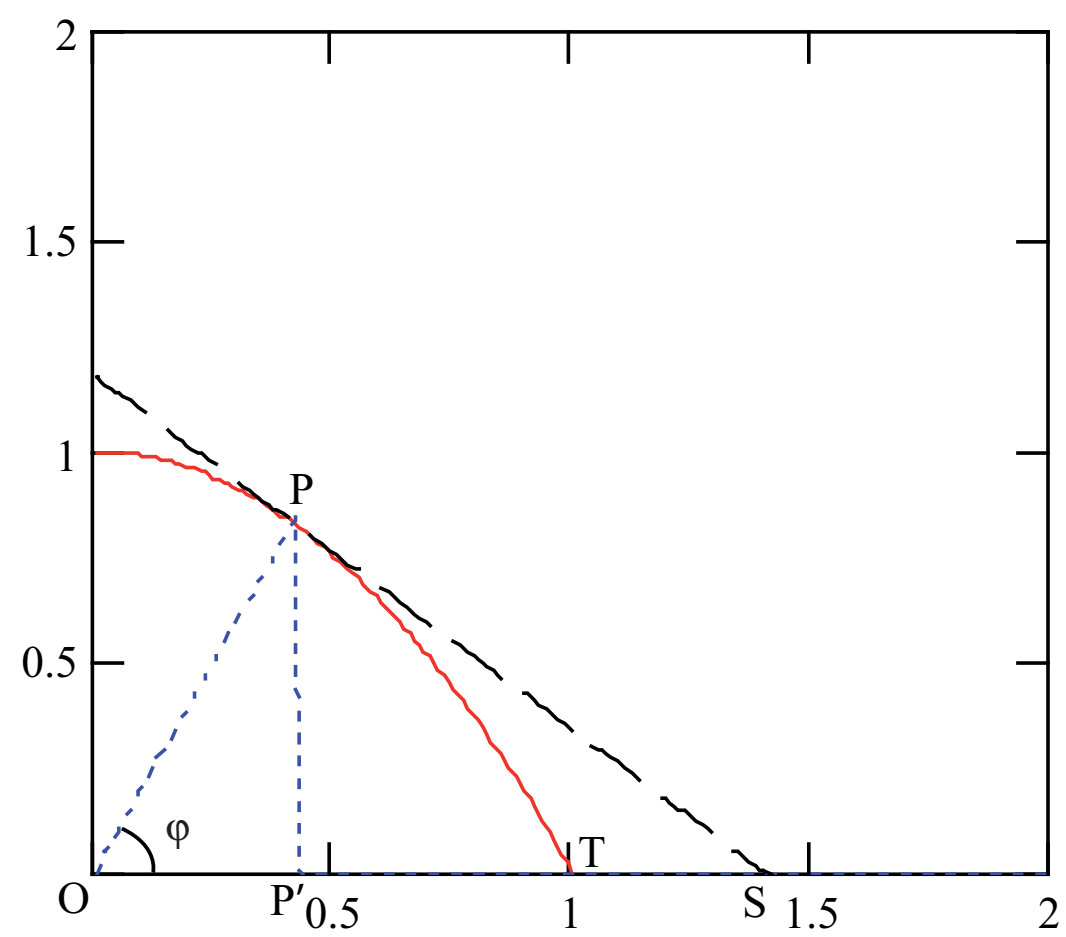

Fig. 3

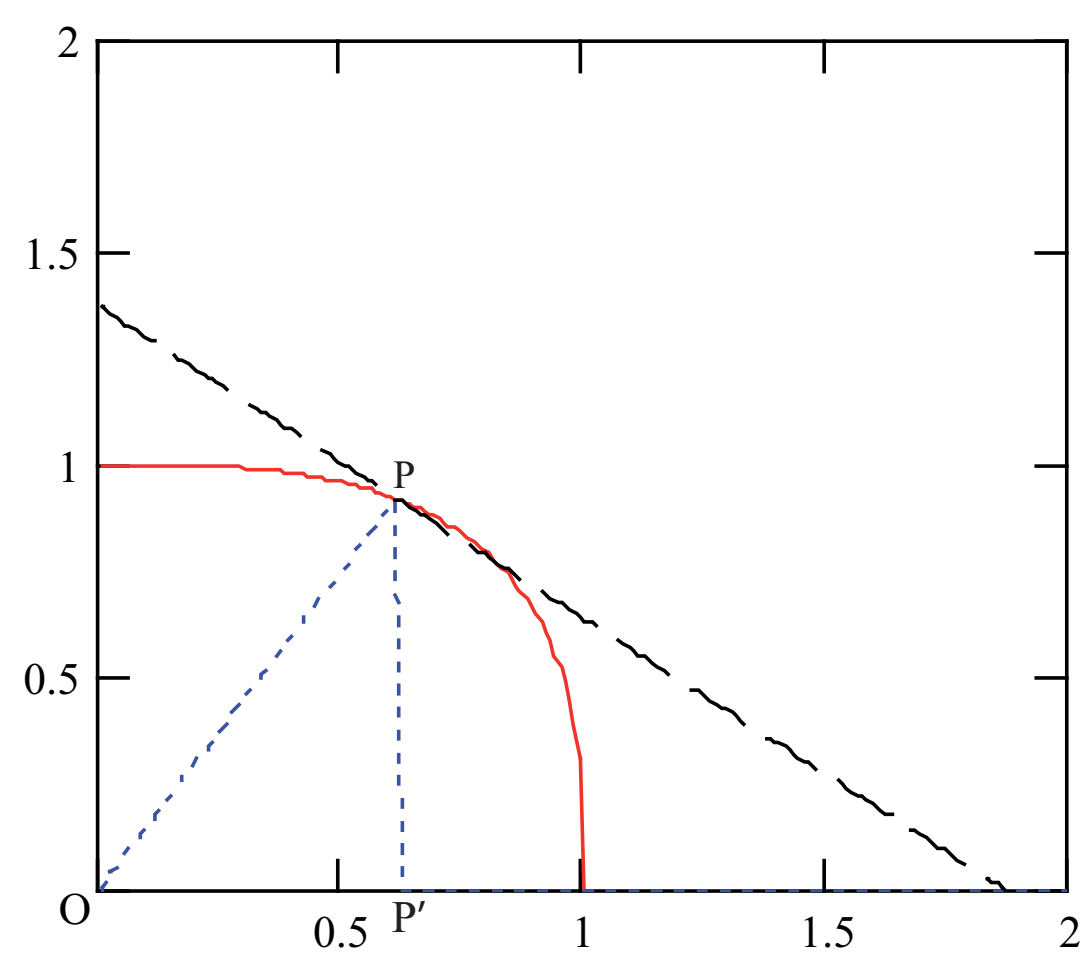

Fig. 4

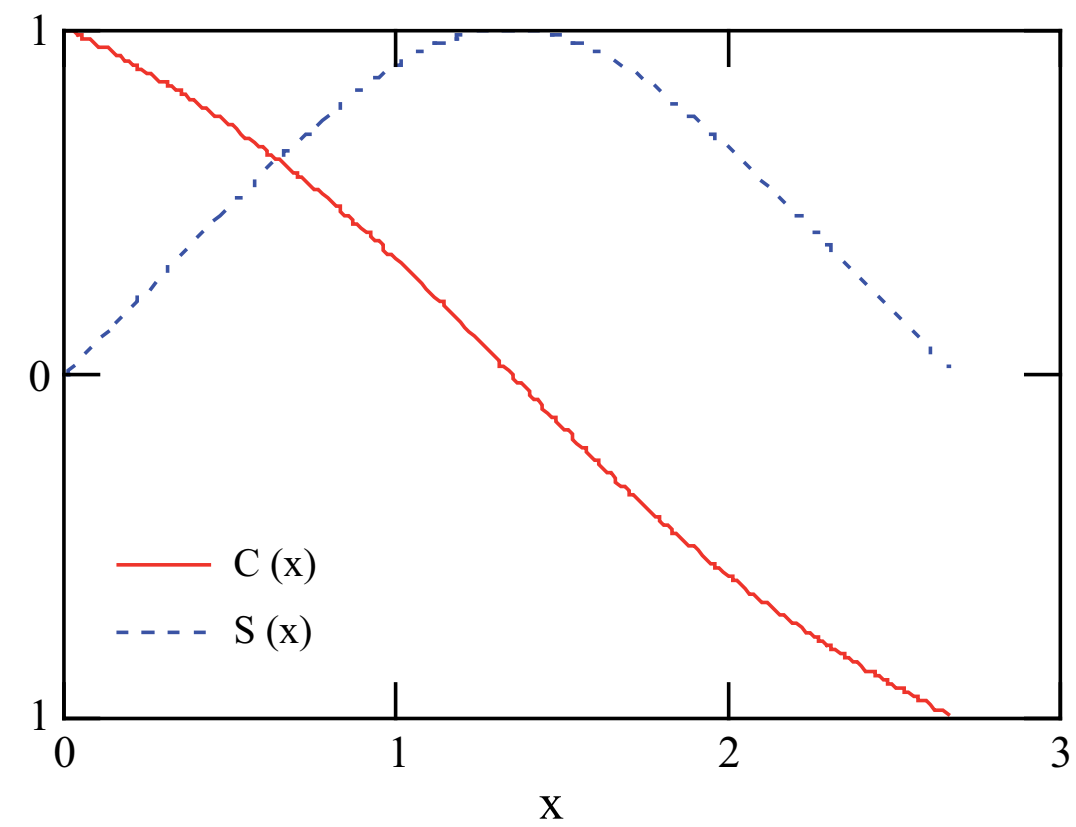

Fig. 5

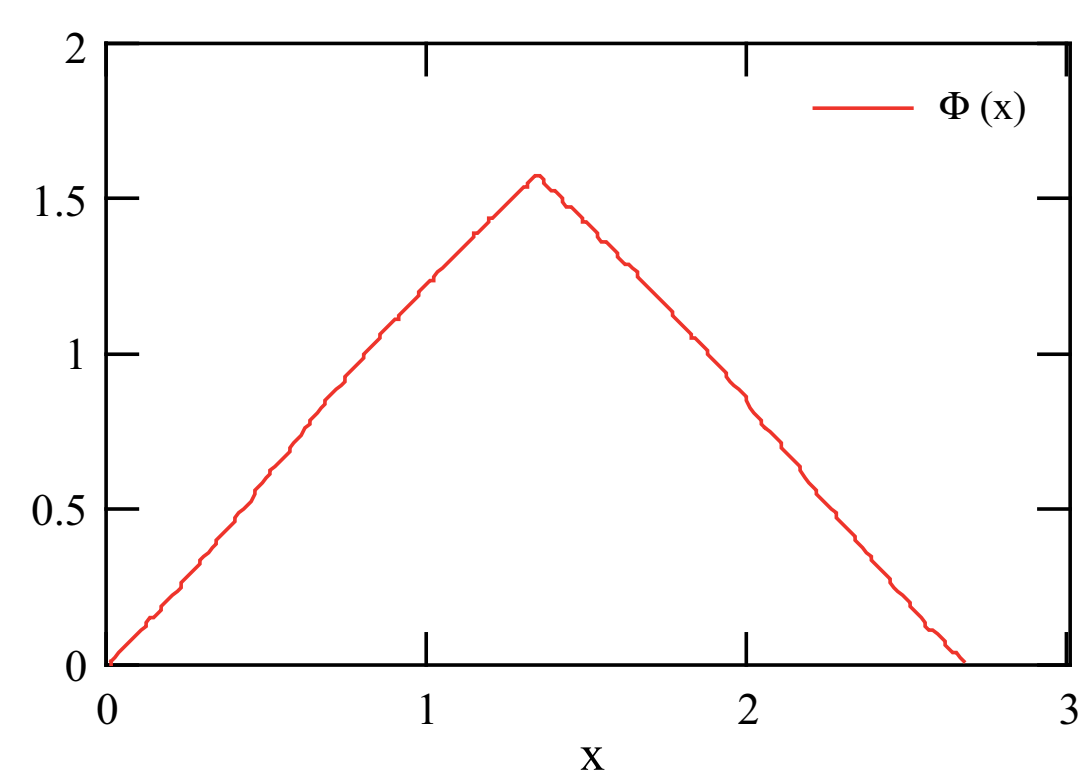

Fig. 6

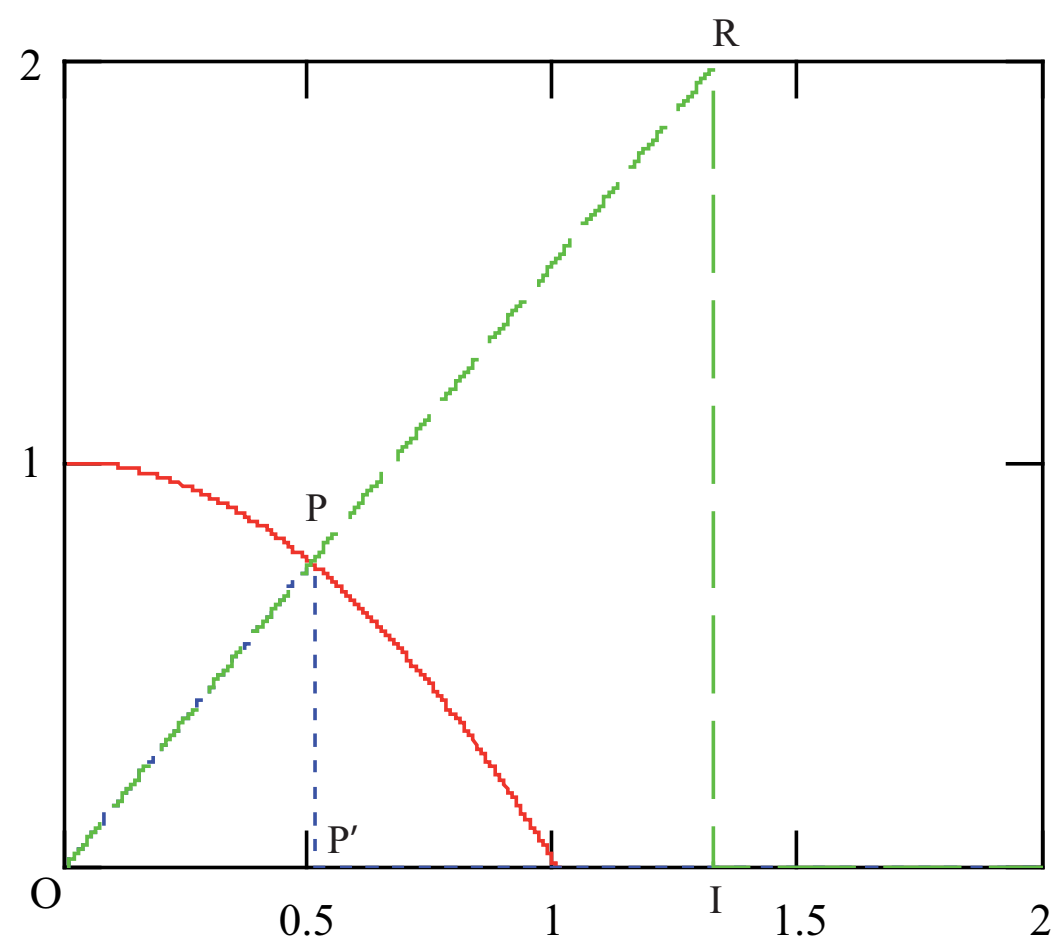

Fig. 7

**FIGURE CAPTIONS**

Fig. 1 Geometrical interpretation of the trigonometric functions $\overline{OP'} = \cos\alpha$, $\overline{PP'} = \sin\alpha$, $\overline{OS} = 1/\cos\alpha$ .

Fig. 2 Geometrical interpretation of the hyperbolic functions. The continuous curve is an arc of hyperbola ($x^2$-$y^2$=1), (), the dash straight line is tangent to the hyperbola at the point P. The sector SOPT represents the area of the “triangle” OPT

$$PP' = \sinh(\Phi), OP' = \cosh(\Phi),$$
$$OS = \frac{1}{\cosh(\Phi)} = \mathrm{sech}(\Phi),\ TT' = \mathrm{tgh}(\Phi),$$
$$\Phi = 2S_{OPT}$$

Fig. 3 Geometrical interpretation of the parabolic functions. The continuous curve is an arc of hyperbola ($y+x^2$=1), the dash straight line is tangent to the hyperbola at the point P. The sector $S_{OPT}$ represents the area of the “triangle” OPT

$$PP' = \sin p(\Phi),\ OP' = \cos p(\Phi),$$
$$\Phi = 2S_{OPT}$$

Fig. 4 Geometrical interpretation of the functions $C(\Phi|3), S(\Phi|3)$. Same as Fig. 3), the continuous curve corresponds to $x^3 + y^3 = 1$

Fig. 5 Parabolic cosine (continuous curve) and parabolic sine (dot curve) vs. $\Phi$

Fig. 6 Parabolic Gudermann function (to avoid discontinuities due to change in sign, we have plotted $tg^{-1}\left[\left|tgp(\Phi)\right|\right]$ vs. $\Phi$

Fig. 7 The elementary trigonometric identity for rectangle triangle. $OI = C, OR = I, RI = c$